\documentstyle[preprint,eqsecnum,aps]{revtex}
\input epsf.sty
\begin{document}
\title{Image charges revisited: a closed form solution}
\author{T.C. Choy}
\address{
Department of Physics and Astronomy,\\
University of Manchester,
Manchester M13, 9PL
United Kingdom
}
\date{\today}
\maketitle
\begin{abstract}
We demonstrate that the corrections to the classical Kelvin image
theory due to finite electron screening length $\lambda$, recently
discussed by Roulet and Saint Jean, Am. J. Phys. {\bf 68}(4) 319,
is amenable to an exact closed form solution in terms of an
integral involving Bessel functions. An improper choice of
boundary conditions is rectified as well, enabling also a complete
solution for all potentials - both {\it inside} and {\it outside}
the metal surface.
\end{abstract}

\section{Introduction}
The theory of image charges over a perfect metal surface has an
early history that dates back to Lord Kelvin in 1848
\cite{Kelvin}. It is a subject covered in most textbooks on
electrostatics \cite{Jackson},\cite{Landau}, as well as an
important topic in modern research \cite{Poladian}, \cite{Choy}.
Hence it is of considerable importance in the undergraduate
curriculum. In a recent article in this journal, Roulet and Saint
Jeans (RSJ) \cite{BRandMSJ} discussed the corrections beyond the
classical Kelvin theory by considering the effect due to a finite
electron screening length $\lambda$ in the metal. This topic has
indeed hardly been addressed in any textbooks, except for more
advanced texts such as Mahan \cite{Mahan}, but only in the more
general theory of electron screening and response functions. A
simplified consideration based on chemical equilibrium such as
employed by RSJ \cite{BRandMSJ} should deepen the understanding of
classical Kelvin image theory as well as stimulate advanced
students to consider the more comprehensive treatment based on
many-body theory and linear response \cite{Mahan}. Unfortunately
the otherwise excellent exposition of RSJ \cite{BRandMSJ} is
plagued by an improper choice of boundary conditions, and by their
inability to obtain a complete solution for the potential in the
metal dictated by a Helmholtz type partial differential equation
(PDE), which as we shall see is in fact separable, see Appendix I
and II. They resorted to a perturbative solution that is not valid
for large screening lengths - which is in fact a trivial limit. In
this paper we shall rectify this and consequently provide the
complete solution for the potentials both {\it inside} and {\it
outside} the metal surface. Naturally our results, being exact
will recover the expected classical theory both in the case of
small screening length ($\lambda \rightarrow 0$), as treated by
RSJ \cite{BRandMSJ}, and the opposite case of large screening
length ($\lambda \rightarrow \infty$), when the metal becomes
ineffective and the classical Laplace potential ensues.

This paper is divided into four sections.  In section I we shall
review the classical Kelvin image theory, solved by the separation
of the Laplace equation in cylindrical co-ordinates. This detail
solution is nowadays not commonly taught in the undergraduate
curriculum \cite{Smythe}. In section II we shall demonstrate that
by using the same technique, the Helmholtz equation (Eq.(22) of
ref \cite{BRandMSJ}),  for the screened potential inside the metal
can also be solved. More importantly we shall discuss the proper
boundary conditions for this problem. The standard conditions
follow from the Maxwell equations \cite{Jackson}: ($div\ {\bf D}=
\rho$ and $curl\ {\bf E}=0$) on the surface edge ($z=0$). However
RSJ imposed an artificial condition based on a {\it strict}
compliance of the {\it final} surface charge density (here denoted
as $\sigma(r)$) with the limiting classical surface charge density
(here denoted as $\sigma_0(r)$). Here we shall argue that the
limiting surface charge density $\sigma_0(r)$ should only be a
consequence of the complete theory and not a precondition. As a
result $\sigma(r)$ should be $\lambda$ dependent, while
$\sigma_0(r)$ is manifestly not. The RSJ choice of boundary
condition, as it will be shown is also amenable to a closed form
solution (see Appendix II), hence their perturbative approach is
unnecessary. It is interesting to note (see Appendix II) that the
exact solution to their problem in fact diverges as $\lambda^2$
for large $\lambda$ (see Appendix II) and thus the perturbation
method is inoperable in principle, as it can never converge to the
exact solution. We believe this essentially non-perturbative
feature is an artifact of their boundary condition which is unlike
the solutions presented here. Moreover, although the departures
between our theory and theirs only become apparent at
$O(\lambda^2)$ for small $\lambda$, their theory further leaves
the outside potentials undetermined and will require additional
assumptions to complete. We shall show that our solution to the
problem is complete and can be expressed in terms of integrals
involving Bessel functions of the type similar to and much studied
in electromagnetic propagation \cite{Kraus},\cite{Stratton},
beginning with Sommerfeld's classic work of 1909
\cite{Sommerfeld}.  In section III shall present the exact
solutions and in section IV we shall discuss the effects on and
hence corrections to the classical image potentials outside the
metal surface, which were omitted by RSJ \cite{BRandMSJ}. Here we
shall also obtain the corrections to the surface charge density as
discussed above. This correction integrates to zero charge as we
shall see (Appendix III). In section V we shall replace the metal
in our theory by a dielectric with $\epsilon>\epsilon_0$.
The reader may like to note that all the results of 
this paper are not new. They have in fact been derived before 
(unknown to the author) by Newns \cite{Newns} and more recently by Kr{\v c}mar
et al \cite{Krumar} (see the end notes).
As a result of our investigations, the conclusions reached by RSJ for
the case of large $\lambda/h>>1$ as in a semiconductor is subtle as
naively the limit $\lambda\rightarrow\infty$ in fact approaches the 
classical results as required for a dielectric. A more careful comparison
will require investigating the case $\lambda >> h$ under the
condition that $z,r >> \lambda$ which is indeed a non-trivial
limit in our theory. More detailed numerical calculations can be found in
ref.\cite{Krumar}.  We shall conclude with a brief discussion of
time-dependent effects, such as the case of an oscillating charge,
by which ingenious experiments based on the classical skin effect
could be used to test our predictions. These experiments could be
designed and possibly be inspiring to an undergraduate class.
\section{Classical Image theory}
In this section we shall adopt the more common convention in which
the physical charge $q$ is introduced above the metal surface at
$z=h>0$ and where the rest of the metal is defined for $z \le 0$,
see Fig 1. This is opposite to the case of RSJ but is similar to
most textbooks such as Jackson \cite{Jackson} and Landau et al
\cite{Landau}. It is well known that Kelvin image theory provides
the solution for the potential outside the metal as:
\begin{equation}
\phi(r,z)={q \over 4\pi\epsilon_0\sqrt{r^2+(h-z)^2}} - {q \over
4\pi\epsilon_0\sqrt{r^2+(h+z)^2}}. \label{eqn1}
\end{equation}
The second term is the so called image potential due to the
fictitious charge $-q$ whose addition enables the boundary
condition $\phi(r,0)=0$ to be satisfied, see Fig 1. This is a
Dirichlet boundary value problem whose solution Eq(\ref{eqn1}) is
justified heuristically in most texts, see for example
\cite{Landau}. It is perhaps useful in a later undergraduate
course to actually verify that Eq(\ref{eqn1}) is indeed a solution
of the Laplace equation. This can be demonstrated in the following
way. The Laplace equation in cylindrical coordinates with
azimuthal symmetry:
\begin{equation}
{\partial^2 \phi^0 \over \partial z^2}+ {1\over r} {\partial \over
\partial r} \Bigl ( r {\partial \over \partial r} \Bigr ) \phi^0 = 0,
\label{eqn2}
\end{equation}
has solutions that are given in terms of Bessel functions \cite{Smythe},
\cite{Panofsky}:
\begin{eqnarray}
\phi^0_>(r,z) &=&  \int_0^\infty f(k) e^{-kz}J_0(kr) dk \qquad
{\rm for} \qquad z>h , \nonumber
\\
\phi^0_<(r,z) &=&  \int_0^\infty g(k) sinh(kz) J_0(kr) dk \qquad
{\rm for} \qquad 0\le z\le h. \label{eqn3}
\end{eqnarray}
The coefficients of these expansions namely $f(k)$ and $g(k)$ are
to be determined by appropriate boundary conditions as we shall
see. Note that the second expansion has been chosen with the
$sinh$ function in order to satisfy the required boundary
condition on the metal surface $z=0$. We have two coefficients and
thus we need two more boundary conditions to determine the
solutions. As is well known from most texts \cite{Jackson},
\cite{Landau}, \cite{Panofsky}, for electrostatics, these are
derived from the two Maxwell equations:
\begin{equation}
curl\ {\bf E} = 0 \qquad {\rm and}\qquad  div\ {\bf D} = \rho,
\label{eqn4}
\end{equation}
which upon integrating around an infinitesimal surface volume
imply the continuity of the potential, say at $z=h$:
\begin{equation}
\phi^0_<(r,h)=\phi^0_>(r,h), \label{eqn5}
\end{equation}
and the discontinuity of the slope:
\begin{equation}
{q \delta (r) \over \epsilon_0} = \Bigl ( {\partial \phi^0_< \over
\partial z} - {\partial \phi^0_> \over \partial z} \Bigr )_{z=h};
\label{eqn6}
\end{equation}
where the charge density $\rho$ consists only of the {\it true}
charge and does not include any induced charges.  The delta
function in Eq.(\ref{eqn6}) is normalized for:
\begin{equation}
2\pi \int_0^\infty r \delta(r) dr = 1.
\label{eqn7}
\end{equation}
Now the continuity condition Eq. (\ref{eqn5}) leads to:
\begin{equation}
f(k) = g(k) e^{kh}sinh(kh), \label{eqn8}
\end{equation}
while the discontinuity condition  Eq. (\ref{eqn6}), with the use
of the orthogonality property of the Bessel functions
\cite{Morse}:
\begin{equation}
\int_0^\infty r J_0(kr) J_0(k^{\prime}r) dr = {1\over
k}\delta(k-k^{\prime}), \label{eqn9}
\end{equation}
leads to:
\begin{equation}
{q\over 2\pi\epsilon_0}=g(k) cosh(kh)+f(k) e^{-kh}. \label{eqn10}
\end{equation}
The straightforward solution of the simultaneous Eq.(\ref{eqn8}) and
Eq.(\ref{eqn10}) leads to:
\begin{eqnarray}
f(k)&=&{q\over 2\pi\epsilon_0}sinh(kh) \qquad {\rm and} \nonumber
\\
g(k)&=&{q\over 2\pi\epsilon_0}e^{-kh}. \label{eqn11}
\end{eqnarray}
That these solutions lead to the classical Kelvin image potential
Eq.(\ref{eqn1}) follows readily from another student exercise in
the form of a mathematical identity for Bessel functions
\cite{Arfken}:
\begin{equation}
{1\over \sqrt{r^2+h^2}}= \int_0^\infty e^{-k|h|}J_0(kr) dk .
\label{eqn12}
\end{equation}
Notice that the final solution depends on the boundary conditions
{\it both} at $z=h$ and at $z=0$, and not just on the latter alone
as the heuristic argument \cite{Landau} seems to suggest. Finally
to conclude this section we shall derive the classical {\it
induced} surface charge density $\sigma_0(r)$. This is simply
obtained from the required boundary condition \cite{Landau}, \cite{Panofsky}:
\begin{equation}
\sigma_0(r) = -\epsilon_0 {\partial \phi^0_> \over \partial z}
{\Big |}_{z=0} = -{q\over 2\pi\epsilon_0} \int_0^\infty e^{-k|h|}
k J_0(kr) dk = -{q\over 2\pi\epsilon_0} {h\over (r^2+h^2)^{3/2}} ,
\label{eqn13}
\end{equation}
a result easily derived using Eq.(\ref{eqn12}). We remind the
reader that this boundary condition is obtained from an
infinitesimal small surface volume integral of the Maxwell
equation:
\begin{equation}
div\ {\bf E} = {\rho_t \over \epsilon_0}, \label{eqn14}
\end{equation}
in the same way as  Eq(\ref{eqn6}) where the total charge density
$\rho_t$ contains {\it all} charges, true and induced
\cite{Panofsky}. In the section IV, as with RSJ
\cite{BRandMSJ}, we shall derive this surface charge density by
integrating the bulk interior charge density $\rho^0(r,z)$ as we
shall see.

\section{Bulk charge density - an exact solution}
The (Thomas-Fermi) screening theory modifications for the potential inside the
metal $\phi_{in}$ is given by the Helmholtz equation \cite{BRandMSJ}:
\begin{equation}
{\partial^2 \phi_{in} \over \partial z^2}+ {1\over r} {\partial
\over \partial r} \Bigl ( r {\partial \over \partial r} \Bigr )
\phi_{in} = {\phi_{in} \over \lambda^2} \qquad {\rm for} \qquad
z<0, \label{Helmholtz}
\end{equation}
where the screening length $\lambda$ is given by:
\begin{equation}
\lambda^2= {\epsilon_0 \over e^2} \Bigl ( {\partial \mu_0 \over \partial n_0} \Bigr )_T.
\label{ScreeningLength}
\end{equation}
This is expressed in terms of a thermodynamic derivative of the
chemical potential with respect to the density.  We note that this
theory for $\lambda$ given by eqn(\ref{Helmholtz}) is in fact
quite general, it should also be valid in the case of a dielectric
or a liquid surface, upon replacing $\lambda$ by the appropriate
Debye-H{\"u}ckel screening length. Significant modifications to
the Thomas-Fermi theory will come about only when $\lambda$
becomes comparable to the Fermi wavelength $\lambda_F$ which we
shall briefly mention in the conclusion. In Appendix I, we shall
show that the solution of the Helmholtz equation
Eq.(\ref{Helmholtz}) is also expressed in terms of Bessel
functions. Hence the complete solution of boundary value problem
specified by Eq.(\ref{eqn2}) and Eq.(\ref{Helmholtz}) is now given
by:
\begin{eqnarray}
\phi_>(r,z) &=&  \int_0^\infty f(k) e^{-kz}J_0(kr) dk \qquad {\rm
for} \qquad z>h , \nonumber
\\
\phi_<(r,z) &=&  \int_0^\infty (g_1(k)e^{kz} + g_2(k) e^{-kz})
 J_0(kr) dk \qquad {\rm for} \qquad 0\le z\le h \qquad {\rm and},
 \nonumber \\
\phi_{in}(r,z) &=& \int_0^\infty g_3(\beta_k) e^{\beta_k z}
{k\over \beta_k} J_0(k r) dk \qquad {\rm for} \qquad z<0 ;
\label{LaplaceHelmholtz}
\end{eqnarray}
where $\beta_k^2 = k^2+1/\lambda^2$, see Appendix I. Note that the
non-vanishing of the potential on the surface $(z=0)$ for general
$\lambda$ requires that we construct solutions for $\phi_<$ that
contains two (in general) unequal coefficients $g_1(k)$ and
$g_2(k)$. We have already discussed the boundary conditions at
$z=h$. The continuity condition Eq.(\ref{eqn5}) leads to:
\begin{equation}
f(k)-g_2(k)= g_1(k)e^{2kh}, \label{Cont1}
\end{equation}
while the discontinuity condition Eq.(\ref{eqn6}) leads to:
\begin{equation}
g_1(k)= {q\over 4\pi\epsilon_0} e^{-kh}. \label{Cont2}
\end{equation}
Two more boundary conditions are required to obtain the solution.
These are in fact obtained in the same way as Eq.(\ref{eqn5}) and
Eq.(\ref{eqn6}), only that now we have continuity of the
potentials and their slopes on the surface $z=0$ in the absence of
true charges on the surface. Thus from:
\begin{equation}
\phi_{in}(r,0)=\phi_<(r,0), \label{Cont3}
\end{equation}
we obtain:
\begin{equation}
g_1(k)+g_2(k)={k\over \beta_k}g_3(\beta_k), \label{gSol1}
\end{equation}
and from:
\begin{equation}
{\partial \phi_{in} \over
\partial z}\Bigr |_{z=0} = {\partial \phi_< \over \partial z}\Bigr |_{z=0},
\label{Cont4}
\end{equation} we obtain:
\begin{equation}
g_1(k)-g_2(k)=g_3(\beta_k). \label{gSol2}
\end{equation}
The problem is now completely specified by equations:
(\ref{Cont1}),(\ref{Cont2}),(\ref{gSol1}) and (\ref{gSol2}). A
simultaneous solution of these equations leads to a complete
solution for all the potentials, hence:
\begin{eqnarray}
\phi_>(r,z) &=&  {q\over 2\pi\epsilon_0} \int_0^\infty \Bigl [{k
cosh(kh)+\beta_k sinh(kh) \over k+\beta_k} \Bigr ] e^{-kz}J_0(kr)
dk \qquad {\rm for} \qquad z>h , \nonumber
\\
\phi_<(r,z) &=& {q\over 2\pi\epsilon_0} \int_0^\infty \Bigl [{k
cosh(kz)+\beta_k sinh(kz) \over k+\beta_k} \Bigr ] e^{-kh}
 J_0(kr) dk \qquad {\rm for} \qquad 0\le z\le h \qquad {\rm and},
 \nonumber \\
\phi_{in}(r,z) &=& {q\over 2\pi\epsilon_0} \int_0^\infty {k
e^{\beta_k z} \over k+\beta_k} e^{-kh} J_0(k r) dk \qquad {\rm
for} \qquad z<0
; \label{All potentials1}
\end{eqnarray}
In the Appendix II, we shall present in the same way the exact
solution for the RSJ problem which consist of Eq.(\ref{Helmholtz})
only and the boundary condition that $\sigma(r)=\sigma_0(r)$,
which however leaves the outside potentials unspecified.

\section{Screening Corrections}
Naturally it would be meaningful to compare these potentials with
the classical image potentials in Eq.(\ref{eqn3}) and
Eq.(\ref{eqn11}). To do this each coefficient in Eq.(\ref{All
potentials1}) is rationalized by multiplying with
$(k-\beta_k)/(k-\beta_k)$ and after some simple algebra, we have:
\begin{equation}
\phi_< = \phi^0_< + \phi^{\lambda} \qquad {\rm and} \qquad \phi_>
= \phi^0_> + \phi^{\lambda}, \label{Potentials2}
\end{equation}
where $\phi^0_<$ and $\phi^0_>$ are the classical potentials given
in section II.  The screening corrections are given in terms of
the $\lambda$ dependent potential:
\begin{equation}
\phi^{\lambda}(r,z) = {q \lambda^2\over 2\pi\epsilon_0}
\int_0^\infty\Bigl(\sqrt{k^2+{1\over \lambda^2}}-k\Bigr )
e^{-k(h+z)} k J_0(kr) dk . \label{QPotential}
\end{equation}
We can easily derive that for small $\lambda$:
\begin{equation}
\phi^{\lambda}(r,z)\Big|_{\lambda\rightarrow 0} \approx {q
\lambda\over 2\pi\epsilon_0} \int_0^\infty e^{-k(h+z)} k J_0(kr)
dk = {q \lambda\over 2\pi\epsilon_0}
{(h+z)\over[r^2+(h+z)^2]^{3/2}} . \label{Limit1}
\end{equation}
For large $\lambda$ we use an expansion of the square
root:$\sqrt{1+1/x^2}\approx1+1/(2x^2)$ to obtain:
\begin{equation}
\phi^{\lambda}(r,z)\Big|_{\lambda\rightarrow \infty} \approx {q
\over 4\pi\epsilon_0} \int_0^\infty e^{-k(h+z)}  J_0(kr) dk = {q
\over 4\pi\epsilon_0} {1\over\sqrt{r^2+(h+z)^2}}, \label{Limit2}
\end{equation}
which cancels the image potential in Eq.(\ref{eqn1}) as required.
Using the same manipulations we obtain the inside potential as:
\begin{equation}
\phi_{in}(r,z) = {q \lambda^2\over 2\pi\epsilon_0}
\int_0^\infty\Bigl(\sqrt{k^2+{1\over \lambda^2}}-k\Bigr )
e^{-kh}e^{z\sqrt{k^2+{1\over \lambda^2}}}k J_0(kr) dk ,
\label{InsidePotential}
\end{equation}
from which the bulk charge density:
\begin{equation}
\rho(r,z)=-{\epsilon_0\over\lambda^2} \phi_{in}(r,z)={-q\over
2\pi} \int_0^\infty\Bigl(\sqrt{k^2+{1\over \lambda^2}}-k\Bigr )
e^{-kh}e^{z\sqrt{k^2+{1\over \lambda^2}}}k J_0(kr) dk
\label{BulkChargeDensity}
\end{equation}
is obtained. In particular we have the limit:
\begin{equation}
\phi_{in}(r,z)\Big|_{\lambda\rightarrow 0} \approx {q \lambda\over
2\pi\epsilon_0} \int_0^\infty e^{-kh} e^{z/\lambda} k J_0(kr) dk =
{q \lambda\over 2\pi\epsilon_0}
{h\over[r^2+h^2]^{3/2}}e^{z/\lambda} \qquad (z<0) ,
\label{PhiLLimit}
\end{equation}
in agreement with RSJ \cite{BRandMSJ}.  However it is easy to show
that the higher order terms differ, see Appendix II. Moreover
unlike RSJ, we have no difficulty with the limit of large
$\lambda$ which again by an expansion of the square root leads to:
\begin{equation}
\phi_{in}(r,z)\Big|_{\lambda\rightarrow \infty} \approx {q \over
4\pi\epsilon_0} \int_0^\infty e^{-k(h-z)} J_0(kr) dk = {q\over
4\pi\epsilon_0} {1\over\sqrt{r^2+(h-z)^2}} \qquad (z<0) ,
\label{PhiInLimit}
\end{equation}
which is of course the correct Laplace potential for the case when
the metal is ineffective, since the Helmholtz equation
Eq(\ref{Helmholtz}) now reduces to the Laplace equation. As
promised in section I, we shall evaluate the surface charge
density:
\begin{equation}
\sigma(r)=\int_{-\infty}^0 \rho(r,z)
dz=\sigma_0(r)+\sigma^{\lambda}(r), \label{SurfaceChargeDensity1}
\end{equation}
where
\begin{equation}
\sigma^{\lambda}(r)={q\over 2\pi}\int_0^\infty {e^{-kh}\over
\sqrt{k^2+1/\lambda^2}} k^2 J_0(kr) dk,
\label{SurfaceChargeDensity2}
\end{equation}
which is the new term in our theory.  This quantity integrates to
zero charge, (see Appendix III), thus the total charge remains
$-q$ as before. Hence the form $\sigma_0$ is not the only (unique)
surface charge density that integrates to a total charge of $-q$.
We note that while the limit $\lambda\rightarrow\infty$ appears
straightforward, this is deceptive. Attempts to calculate the next
order i.e. $O(1/\lambda^4)$ corrections lead to divergent
integrals that require careful treatment. We shall not discuss
this exercise here. Suffice to say the corrections to the
classical potentials due to $\phi^{\lambda}$ and $\phi_{in}$ are
to smear the image charge from a point charge to a charge
distribution whose weight vanishes as $\lambda$ increases. This
charge distribution can be derived from the results presented here
and has been worked out by Newns \cite{Newns} in terms of an order
two Bessel function.

\section{Semiconductor surface}
The extension of our theory to a semiconductor surface, in which 
$\lambda$ is finite, is now rather
straightforward. The only significant modification is that the
boundary condition Eq.(\ref{Cont4}) is now replaced by:
\begin{equation}
\epsilon {\partial \phi_{in} \over
\partial z}\Bigr |_{z=0} = \epsilon_0{\partial \phi_< \over \partial z}\Bigr |_{z=0},
\label{Cont4D}
\end{equation}
where $\epsilon>\epsilon_0$ is the dielectric constant of the
material. We need not repeat the details here and present merely
the solution for $\phi_{in}$ which now takes the form:
\begin{equation}
\phi_{in}(r,z) = {q\over 2\pi\epsilon_0} \int_0^\infty {k
e^{\beta_k z} \over k+{\tilde \epsilon}\beta_k} e^{-kh} J_0(k r)
dk \qquad {\rm for} \qquad z<0, \label{PhiInD}
\end{equation}
where ${\tilde\epsilon}=\epsilon/\epsilon_0$.  The integral can be
analyzed in both limits as usual. For small $\lambda$ we have:
\begin{equation}
\phi_{in}(r,z)\Big|_{\lambda\rightarrow 0} \approx  = {q
\lambda\over 2\pi{\tilde\epsilon}\epsilon_0}
{h\over[r^2+h^2]^{3/2}}e^{z/\lambda} \qquad (z<0) ,
\label{PhiLDLimit}
\end{equation}
which is analogous with the metallic case. For large $\lambda$
however:
\begin{equation}
\phi_{in}(r,z)\Big|_{\lambda\rightarrow \infty} \approx {q \over
2\pi(1+{\tilde\epsilon})\epsilon_0} \int_0^\infty e^{-kh} e^{kz}
J_0(kr) dk = {q \over 2\pi(1+{\tilde\epsilon})\epsilon_0}
{1\over\sqrt{r^2+(h-z)^2}} \qquad (z<0) , \label{PhiLDLimit2}
\end{equation}
which is the classical result. Recall that the potential inside a
dielectric is equivalent to an image charge ${\tilde q}$ given by
\cite{Panofsky}:
\begin{equation}
{\tilde q} = {2 {\tilde\epsilon}q \over (1+{\tilde\epsilon})}
 , \label{ImageD}
\end{equation}
replacing the real charge $q$ at the point $h$ outside the
surface. Hence for large $\lambda>>h$, classical theory is
recovered and as a matter of fact it is for $\lambda<<h$, as shown
in Eq.(\ref{PhiLDLimit}) that screening creates departures from
classical theory. Thus for dielectrics it is the case of small and
finite $\lambda$ that leads to corrections to the Kelvin theory.
Finally we shall briefly mention the case of an oscillating
charge. The results here being static will lead to a zero
frequency contribution to the classical finite frequency skin
effect. The propagation of electromagnetic waves in the substrate
will now be dictated by a complex wave vector \cite{ChoyBook}:
\begin{equation}
k^2= {\tilde\epsilon}\omega^2/c^2+ 2 i/\delta^2 - 1/\lambda^2
\label{Complexk}
\end{equation}
to which we have added the screening length $\lambda$. At low
frequencies, \cite{ChoyBook}, the skin depth $\delta$ which varies
as the inverse square root of the frequency dominates. Hence by
extrapolating skin depth measurements to zero frequency, the
results presented in this paper may be detectable \cite{Ref1}.
This could be a novel undergraduate experiment for physics and
engineeering classes.

\section{Conclusion}
In conclusion we have presented an exact solution for the
corrections to the classical Kelvin image theory of electrostatics
previously discussed by RSJ \cite{BRandMSJ} in this journal. Some
inadequacies in their analyses are rectified and a complete
solution for all the potentials are obtained in closed form. We
found that some care needs to be exercised with regard to
statements about non-classical behaviour which are only valid in
the case of {\it finite} $\lambda$, whereas for small and large
$\lambda$ we have shown that the theory reduce to the standard
textbook analysis. Nevertheless our theory shows that the
non-vanishing potentials $\phi^{\lambda}$ and $\phi_{in}$, for any
finite $\lambda>0$ in the case when the charge is right on the
metal surface i.e. for $h=0$ as opposed to the classical theory,
have an essential role for surface chemistry that should be noted
in all textbooks \cite{Zangwill}.

\subsection{Notes added:}

After this work was completed we were grateful to be advised by
Gabriel Barton and Bernard Roulet that the results of our paper, with the
exception of the proofs in the appendices, have in fact been derived
some thirty years ago by Dennis Newns \cite{Newns}
apart from minor differences in
our definition of the induced potentials. 
Also unknown to the author, the results have also been rederived
recently by Kr{\v c}mar et al \cite{Krumar}. The results of this paper 
are in agreement with Newns and Kr{\v c}mar et al 
The reader might be interested to know that the latter authors have also
considered the case when the charge q becomes immersed {\it inside} the 
material, using extensions of the methods presented here. 
Neither Newns,  Kr{\v c}mar et al, nor the
present author have considered the limit when $\lambda_F$ becomes
significant. The author is further indebted to  Marshall
Stoneham for pointing out that the solution in the limit of large
$\lambda_F>>\lambda_{TF}$ has been considered by John Willis
\cite{Stoneham}, {\it only within a static charge approximation}.
This identified the significance of Fermi surface effects (omitted
in this work) which lead to Friedel type oscillations in the
density near the surface. No doubt a more sophisticated theory
such as one using a density functional approach will be needed to
treat all the various issues.

\subsection{Appendix I - solution of the Helmholtz equation}
The Helmholtz equation Eq.(\ref{Helmholtz})
in azimuth symmetric cylindrical coordinates can be easily
separated \cite{Arfken} by the ansatz $\phi(r,z)=f(r)g(z)$, such
that:
\begin{equation}
{d^2 f \over dz^2}=k^2 f \label{Feqn}
\end{equation}
which has solutions: $f(z)=e^{kz}$, where $k^2$ is the separation
constant. Note that we have $z<0$ so that the other solution
$f(z)=e^{-kz}$ is exponentially increasing and cannot be admitted
by the boundary condition for $z\rightarrow -\infty$. The radial
equation now takes the form:
\begin{equation}
{d^2 g \over dz^2}+{1\over r}{d g \over dz}+\alpha_k g=0,
\label{Geqn}
\end{equation}
where: $\alpha^2_k=(k^2-{1\over\lambda^2})$. For $\alpha^2_k>0$
this is the differential equation for Bessel functions of order
zero. For $\alpha_k=0$ the solution goes as $\log r$ which is
inadmissible as is the case where $\alpha^2_k<0$ since Bessel
functions of imaginary arguments namely $I_0(r)$ and $K_0(r)$ are
also inadmissible. Thus the general solution of
Eq.(\ref{Helmholtz}) is a superposition of solutions using an
expansion coefficient $g_3(k)$ given by:
\begin{equation}
\phi_{in}(r,z) = \int_{1/\lambda}^\infty g_3(k) e^{k z}
J_0(\alpha_k r) dk \qquad {\rm for} \qquad z<0 . \label{HeqnSoln}
\end{equation}
By a straightforward change of variable $x=\alpha_k$ we easily
obtain the solution given by the last of
Eqn.(\ref{LaplaceHelmholtz}).
\subsection{Appendix II - exact solution of the RSJ problem}
The solution for the RSJ boundary value problem can be obtained
using the results of Appendix I. The RSJ potential and hence the
bulk charge density is given by:
\begin{equation}
\rho(r,z) = -{\epsilon_0\over\lambda^2} \int_0^\infty g_3(\beta_k)
e^{\beta_k z} {k\over\beta_k}J_0(k r) dk \qquad {\rm for} \qquad
z<0 . \label{RSJSoln1}
\end{equation}
The RSJ boundary condition requires that the integral:
\begin{eqnarray}
\int_{-\infty}^0 dz \rho(r,z)&=& -{\epsilon_0\over\lambda^2}
\int_0^\infty g_3(\beta_k) {k\over\beta_k^2}J_0(k r) dk \nonumber
\\ &=&\sigma_0(r) =  -{q\over2\pi}\int_0^{\infty} e^{-kh} k J_0(kr) dk 
\label{RSJSoln2}
\end{eqnarray}
hence we obtain:
\begin{equation}
g_3(\beta_k)= {q\lambda^2\over 2\pi\epsilon_0}\beta_k^2e^{-kh}.
\label{RSJSoln3}
\end{equation}
Note that while this determines the inside potential in closed
form:
\begin{equation}
\phi^{RSJ}_{in}(r,z) = {q\lambda^2\over  2\pi\epsilon_0 }
\int_0^\infty e^{-kh} e^{\beta_k z} \beta_k k J_0(k r) dk ,
\label{RSJSoln4}
\end{equation}
it leaves the outside potentials undetermined since
Eq.(\ref{Cont1}) and Eq.(\ref{Cont2}) now become two equations
with {\it three} unknowns. Moreover Eq.(\ref{RSJSoln4}) is now
divergent as $\lambda^2$ for $\lambda\rightarrow\infty$ and does
not reduce to the classical Laplace potential as required. It is
interesting to compare Eq.(\ref{RSJSoln4}) with
Eq.(\ref{QPotential}). The former is essentially non-perturbative
whereas the latter may be obtainable perturbatively if some care
is exercised in treating the asymptotic integrals, see the remarks
at the end of section IV.  It is unclear and it would be
interesting to speculate if this feature of the RSJ solution has
experimental significance such as when the total surface charge
can be maintained constant, i.e. $\sigma_0(r)=const$. In this case, 
following the Eq.(\ref{RSJSoln2}), we can show that the potential now 
goes as $\lambda e^{-z/\lambda}$.
Nevertheless it is interesting to compare the small $\lambda$
expansion for the two solutions. We note that the difference
between  Eq.(\ref{RSJSoln4})and Eq.(\ref{InsidePotential}) is a
negative term:
\begin{equation}
\phi_{in}(r,z) = {q\lambda^2\over  2\pi\epsilon_0 } \int_0^\infty
e^{-kh} e^{\beta_k z} (\beta_k-k) k J_0(k r) dk .
\label{SolnChoy1}
\end{equation}
For small $\lambda$ we change variable to $k=x/h$ and carry out
the expansions for the square root and exponential terms via:
\begin{equation}
\sqrt{1+({x\lambda\over h})^2}\approx 1 + {1\over
2}({x\lambda\over h})^2+\dots  \label{SolnChoy2} \qquad {\rm and}
\end{equation}
\begin{equation}
e^{{z\over \lambda}(\sqrt{1+({x\lambda\over h})^2})}\approx
e^{z\over \lambda}( 1 + {1\over 2}{z\lambda x^2\over h^2}+\dots)
\label{SolnChoy3}.
\end{equation}
Inserting these into the integrands and evaluating the integrals
we have the limiting expressions to $O(({\lambda\over h})^2)$ as:
\begin{equation}
\phi^{RSJ}_{in}(r,z) \approx {q\lambda e^{z/\lambda}\over
2\pi\epsilon_0}{h\over (h^2+r^2)^{3/2}}\Bigl [1+ {3\lambda\over
2}(z+\lambda){(2h^2-3r^2)\over (h^2+r^2)^2}+\dots\Bigr ] ,
\label{RSJAsymt}
\end{equation}
in agreement with RSJ \cite{BRandMSJ} while our solution has the
expansion:
\begin{equation}
\phi_{in}(r,z) \approx {q\lambda e^{z/\lambda}\over
2\pi\epsilon_0}{h\over (h^2+r^2)^{3/2}}\Bigl [1+ {3\lambda\over
2}(z+\lambda){(2h^2-3r^2)\over (h^2+r^2)^2}-{\lambda\over
h}{(2h^2-r^2)\over (h^2+r^2)} \dots\Bigr ] , \label{ChoyAsymt}
\end{equation}
which differs from the latter at $O(({\lambda\over h})^2)$ with an
additional term .
\subsection{Appendix III - proof that the integral of
$\sigma^\lambda(r)$ vanishes} The proof is straightforward and it
makes use of a simple trick involving a well known property of the
Bessel functions. We need to obtain, denoting $1/\lambda$ by
$\alpha$, the integral:
\begin{eqnarray}
Q(\alpha) &=& q \int_0^\infty r dr \int_0^\infty dk {e^{-kh}\over
\sqrt{k^2+\alpha^2}} k^2  J_0(k r) \nonumber \\ &=&
lim_{\beta\rightarrow 0} \quad q \int_0^\infty r dr \int_0^\infty
dk {e^{-kh}\over \sqrt{k^2+\alpha^2}} k^2 J_0(\beta r) J_0(k r),
\label{Choytrick1}
\end{eqnarray}
which follows from the property that $ lim_{x\rightarrow 0}\quad
J_0(x) = 1$. Making use of the orthogonality property
Eq.(\ref{eqn9}) we now have:
\begin{eqnarray}
Q(\alpha) &=& lim_{\beta\rightarrow 0}\quad q \int_0^\infty dk
{e^{-kh}\over \sqrt{k^2+\alpha^2}} {k^2\over \beta}
\delta(\beta-k) \nonumber \\
  &=& lim_{\beta\rightarrow 0}\quad q {\beta e^{-\beta h} \over
  \sqrt{\beta^2 + \alpha^2}} \nonumber \\
  & = & 0 \qquad {\rm for} \quad  \alpha >0 \nonumber \\
  & = & q \qquad {\rm for}\quad  \alpha =0
\label{Choytrick2}
\end{eqnarray}
respectively. Although $Q(\alpha)$ is a discontinuous function at $\alpha=0$,
the interchange of the limit with the integrals, while not rigorous, may be 
justified by the convergence of the integrals in Eq.(\ref{Choytrick1}) for all
$\alpha \ge 0$ \cite{Arfken}.

\begin{figure}[h]
\begin{center}
\mbox{
\epsfxsize=13.3cm
\epsfbox{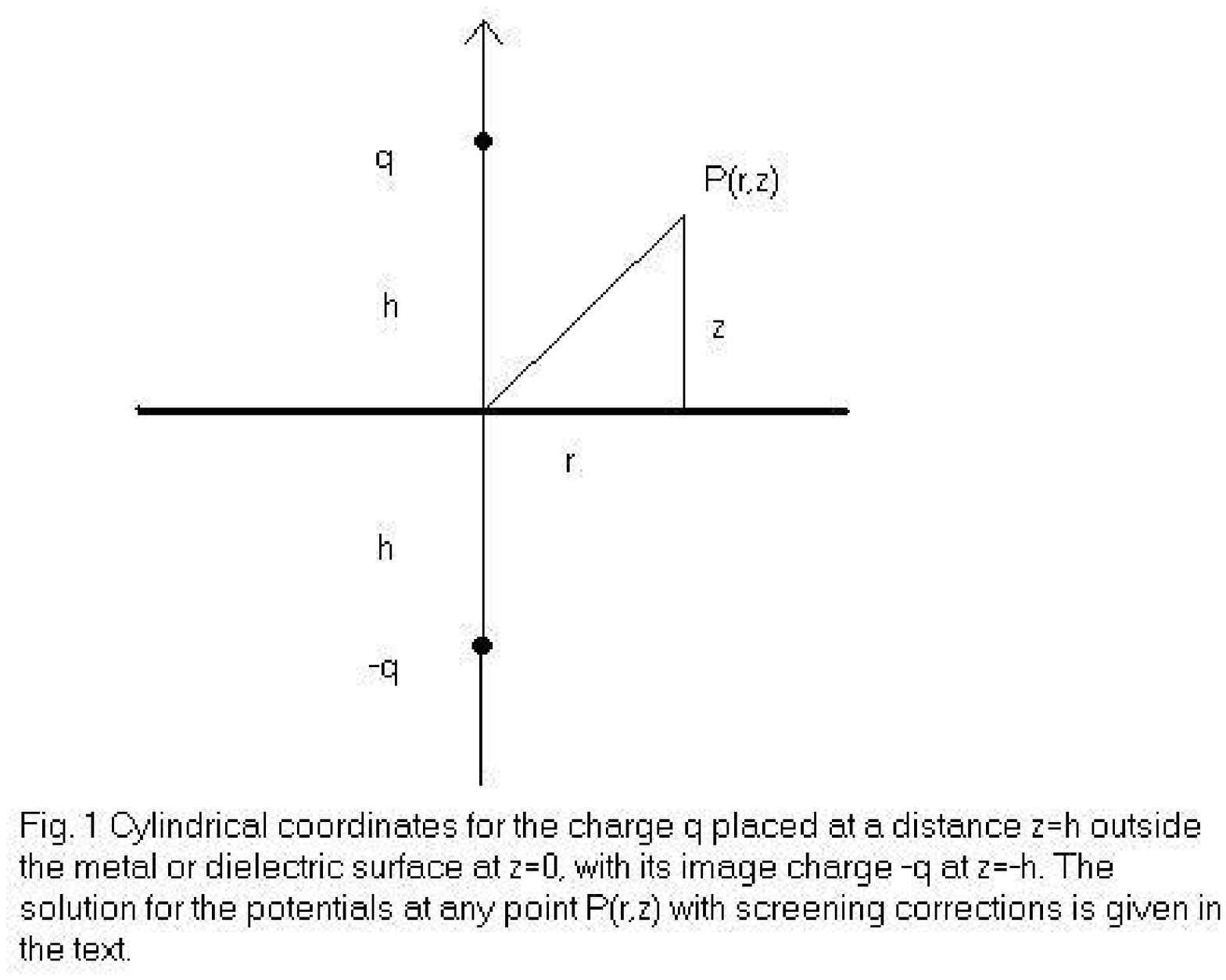}
}
\end{center}
\end{figure}


\begin{references}
\bibitem{Ref1}However note that the equilibrium theory considered so far for
 the screening may need modifications if the frequency is so high as to exceed
a
 typical inverse relaxation time $\tau$ for the electrons which is
usually in the infrared region \cite{Landau}.
\end{references}
\end{document}